\documentclass[floatsintext]{apa7}

\usepackage{hyperref}
\hypersetup{
    colorlinks=true,
    linkcolor=blue,
    filecolor=magenta,      
    urlcolor=cyan,
    allcolors=magenta,
}
\usepackage[american]{babel}

\usepackage{csquotes}
\usepackage[style=apa,sortcites=true,sorting=nyt,backend=biber]{biblatex}
\DeclareLanguageMapping{american}{american-apa}
\title{History Of Rigor: A Review Of 20\textsuperscript{th} Century Science Education}
\shorttitle{History Of Rigor: Science Ed}

\author{Jason Garver}
\affiliation{University Of Minnesota}

\leftheader{Garver}

\abstract{``Rigor'' is an often sought after but ill-defined concept in education. This work reviews several models of rigor from current literature before proposing a tool which is used to analyze science education throughout history. The 20\textsuperscript{th} century science education in the United States was subject to changing sociopolitical motivations about the use of science both in general and for students. These factors as well as developments in theory of learning and broad education reforms had changing affects on the level of rigor in science education. This work analyzes the theoretical level of rigor of science education in the US based on two main motivating factors for science education; science as a social endeavor and science as a discipline, throughout the 20\textsuperscript{th} century.}

\begin{document}
\maketitle

\section{Introduction}
Advancements in both science and education have served as the catalyst for several reforms in science education through the last century. As nationally standardized curricula are a relatively new concept, historically the US sociopolitical climate informed the motivations in science education; what is important to teach, and how to teach it. These factors not only drove what specific science content students learned in secondary school but also the rigor in which students engaged in these topics. Thus, the \textit{``why''} of science education throughout the 20\textsuperscript{th} century can be as informative as textbooks or curricular materials to paint a picture how rigor has changed in science education throughout history. This can serve as a critical background to understand changes in US science education as we move into an increasingly globalized economy with changing political priorities, and more technical advancements in fields such a renewable energy and quantum computing. 

\subsection{Purpose \& Scope}
This work will first provide a \textit{tool} which will be useful in analyzing the rigor of secondary science education throughout history given differences in educational philosophy. This will be used to explore the ways in which factors such as international conflicts, changing views of science, and social reforms outside of education impact the level of rigor for secondary science students. 
\\

The scope of this work will be limited to two main eras: The Space Race of the mid 1940's to late 1960's, then the post Space Race era extending through the '70s, '80s, and '90s. Finally, an overview will be given of modern science education as it pertains the the Common Core Curriculum and more of the reforms of the 2000's. 

\section{Defining Rigor}
Along with shifts in educational priorities, so has the concept of ``rigor'' changed throughout the last hundred years. The word ``rigor'' itself is not a well defined concept today in education and it was even less so throughout history \parencite{Priem}. In a historical context there were many terms for what we think of as ``rigor''; ``scientific Literacy'', ``academically demanding'', ``cognitively demanding'', etc \parencite{Wyse}. Many definitions of rigor today are centered around several main themes; high quality and high status courses and material, critical and creative thinking, and grading practices which reflect those qualities \parencite{Bower}. Given the complexity and interconnections of what rigor means, it is not the goal of this work to \textit{define} rigor in one way, as that definition can take many forms depending on context. Instead, a tool which can be used to analyze rigor in science education through history will be used for comparative analysis for this work. 

\subsection{A Brief Overview Of Rigor Today}
What exactly educators mean by ``rigor'' depends largely on context. One can define rigor in terms of what students are \textit{doing}, the types of assessments they are completing, by collaboration in curriculum development or by ideas from educational psychology.  Furthermore, it has been realized in recent years that rigor is not a binary measure, but both a continuum and multi-dimensional \parencite{Paige}. This section contains a very short overview of some common rigor frameworks today.
\\

One possible definition of rigor suggested was suggested by \textcite{Daggett}, which uses Bloom's Taxonomy and the level of real-world application students engage in. This framework is useful in defining the rigor for educational tasks, but historically science education did not always include the same sorts of task and for this work the author wants to account for that. From a student's perspective, the view of rigorous education takes a different form. When surveyed, introductory biology students at a small liberal arts collage described \textit{workload} as a major factor for a course being ``rigorous'', whereas students taking upper level classes at the same institution characterized rigor as high cognitive demand \parencite{Wyse}. Taking an educator's perspective, \textcite{Oconnor} provides a rigor rubric which splits rigor into three aspects; instruction, curriculum, and assessment.

\subsection{Operational Definition Of Rigor}
As stated above, it is not a goal of this work to create a standard definition of rigor, but rather a framework which can be used to compare different historical eras. To accomplish this, the author thought it best to consider the following features; transfer-ability across different eras of educational philosophy and relativity in comparisons. Two main ideas from current literature are combined to create a four quadrant matrix\footnote{Similar to what \textcite{Daggett} produced using Bloom's Taxonomy}. Cognitive Demand which follows from \textcite{Wyse} and \textcite{Hess} and content depth. The latter, ``content depth'' can be thought of roughly as the amount of content knowledge students are expected to engage in. 
\\

An important distinction to make here is that content depth is not simply ``more content'', but rather generalizations, multiple perspectives and connections between different areas of content knowledge. Cognitive demand is roughly described as the relative ``difficulty'' of learning, compared to what a student can already do. Cognitively demanding tasks are those which require analysis, synthesis, application and generalization by the student which was the distinction made by \textcite{Wyse}.
\\

The importance for this work is that a framework to describe rigor is relative. When comparing a first grade classroom and an eleventh grade classroom, there will be a clear discrepancy in the depth of content and at what level of cognitive demand students are asked to engage in. This will also be the case historically, which is why this framework does not make absolute or quantifiable claims. 

\begin{figure}[!htbp]
    \centering
    \includegraphics[scale=.35]{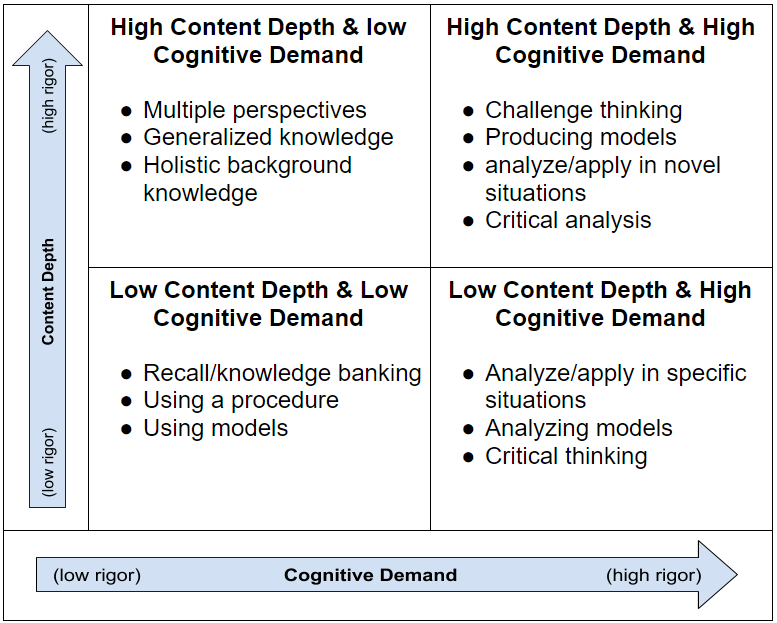}
    \caption{Combined rigor framework based on \textcite{Wyse}, \textcite{Oconnor}}
    \label{fig:Figure1}
\end{figure}

The framework presented in figure \ref{fig:Figure1} is one possible way to characterize rigor for historical comparison. Indicators of rigor are listed within each quadrant of the framework but are non-exhaustive and may overlap, as rigor is a continuum. An important feature to be emphasized here is that increased rigor does not only depend on cognitive demand or content depth but instead each contribute to rigor\footnote{An analogy to this is that of the metric in flat euclidean space: $z= \sqrt{x^2 + y^2}$, where the parts themselves are each smaller than the whole. Rigor can be thought of as ``$z$'' where ``$x$'' and ``$y$'' are content depth and cognitive demand.}, but in different ways. One can imagine a rigorous course which asks students to analyze and produce models based on a relatively superficial level of concepts, or conversely access more complex connections within content at a lower cognitive level.  

\section{Science As A Social Endeavor}
This work will break history into two themes present within the literature, the first of which is ``science as a social endeavor''. These themes are non-exclusive and exhaustive, but serve as a guide to analysis. Science as a social endeavor is categorized in this work as a focus of science education as it applies to issues in society and the perceived needs of students as members of society.

\subsection{Early 1900s: The Scientific Citizen}
The turn of the 20\textsuperscript{th} century was an era of reform in education with works by Dewey and others; science education was no different. With these reforms came an agreement from the Commission on the Reorganization of Secondary Education that science should be relevant and useful to student's lives \parencite{NEA1918}. However, scientists that had a roles in advocating for science in schools thought that science education should include not only practical skills but also to value thinking skills. Though the application of science was a clear focus with the chair of Commission on the Reorganization of Secondary Education, Clarence Kingsley, suggested that science curriculum should be organized in such a way that it supports application to everyday life, rather than scientific endeavors itself \parencite{NEA1918}. This was the challenge throughout the first three decades of the 20\textsuperscript{th} century, where applicability to everyday life brought the fear of losing science as a way of thinking.

\subsection{1970s-1990s: Science As A Tool For Society}
Science education as a need for society popped back up in the 1970's as a way to analyze social issues and make informed decisions about a world that was experiencing a technological boom. It was stressed that understanding the interconnected nature of science, technology, and society was either as important, or more important than science content itself. \textcite{Hurd} even went as far to state that the social context of science was the only appropriate way to organize the educational discipline. The individual needs of students were also an important focus for science education during this time which drove a more content-light but applicable approach \parencite{Yager}.
\\

During the 1980's there was a push to infuse science in society into science education, via the Science-Technology-Society position of the NSTA. The goal of science education during this time was to produce young adults who could analyze science issues in society such that they could make informed decisions about them \parencite{Ramsey}. The STS featured elements of what we call ``critical thinking'' now; deconstructing and identifying science concepts in social issues and analyzing or developing scientific solutions to these issues. Though this was not without criticism; \textcite{Kromhout} argued that for all of the higher order thinking involved in STS, it was lacking in much of the foundational science content and the structure of science as a discipline. 

\subsection{Rigor Of Science As A Social Endeavor}
Using the framework of figure \ref{fig:Figure1}, several commonalities of these time periods emerge. There was a focus on science as it applies to students lives, and also the interplay of science and its role in society with an emphasis on either analyzing or producing solutions to social issues. The drawback of this focus was that some of the ``science'' was lost, or at very least took a back seat to the social application. 

\begin{figure}[!htbp]
    \centering
    \includegraphics[scale=.35]{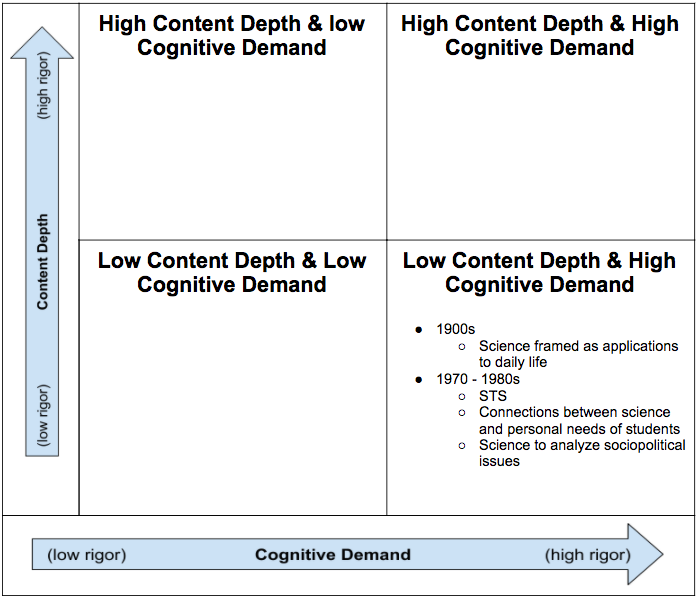}
    \caption{}
    \label{fig:Figure2}
\end{figure}

As shown in figure \ref{fig:Figure2}, application of science to society can be a highly cognitively demanding task, as it requires critical thinking, and some degree of abstract modeling/analysis. As a two dimensional measurement, science as a social endeavor falls short of what could be viewed as \textit{maximum rigor}. As put by \textcite{Kromhout}, the lack of complex connections between content prevents students from challenging current ideas or engaging in the creative process during production of models or application of science to novel situations.

\section{Historical Motivation: Science As A Discipline}
To contrast the last section, the Cold War era saw a rise in concerns of national security which extended to science and technology especially pertaining to rocketry and the successful Russian space probe Sputnik. 

\subsection{Late 1950s: Sputnik Era}
During the 1950's, the united states and its allies were locked in a Cold War with the Soviet Union, the hallmarks of which were technological advancements in nuclear physics, rocketry and space flight. It was this fear of Soviet dominance that gave the US a renewed sense that science was a strategic endeavor. In 1958 the Rockefeller Brothers Fund issued a series of reports, one of which described the state of technological advancement in the US and a call for highly trained individuals to participate in US science \parencite{Rockefeller}. Along with the focus of preparing young people to assist in the advancement of technology, there was a plea for the general public to become sympathetic to science, which required a degree of understanding science itself. This was termed ``scientific literacy'', and was framed as a necessary component for the average person to carry out their civic duty \parencite{Hurd1958}.

\subsection{1960s: Building Scientists}
The natural progression from the anxieties of the 1950's was the space-race of the 1960's. The United States was not only locked in an arms race with the Soviet Union, but also in a race to the moon as a show of strength in technological and scientific advancement. This brought about a more holistic view of science education, where students not only needed deep content knowledge, but also familiarity with the methods and process of science \parencite{Carlton}. This was in part because of the same motivation of the previous decade; the public needed to understand the new advancements being made in science. Science education also needed to support the more elite students who may become what was a precious resource: new scientists. 

\subsection{Rigor Of Science As A Discipline}
Science education throughout the 1950's and 1960's was driven in large part by the Cold War and fears of national security. Where the decades before and after this era focused on science as it applied to students lives, the Cold War era drove the need for greater content knowledge and for students to engage in the process of science as a scientist would.

\begin{figure}[!htbp]
    \centering
    \includegraphics[scale=.35]{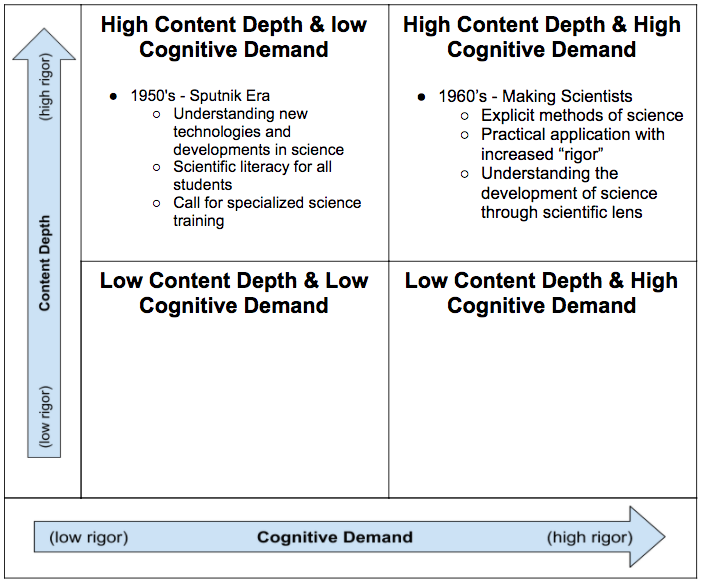}
    \caption{}
    \label{fig:Figure3}
\end{figure}

Pulling together the focus on science knowledge and the scientific process (scientific literacy) from the Cold War era, figure \ref{fig:Figure3} shows that this time period extended across the top of the rigor matrix. This is was born out of the call for a greater understanding of science content in the 1950's as it related to national security, then from the push to the moon during the 1960's. The sociopolitical attitudes of this time period did in fact push science education toward the combination of deep content knowledge and high cognitive demand. One can see that the Cold War era tended toward a higher level of rigor than the decades directly preceding or following it.

\section{Implications For Modern Science Education}
The modern era of science education began in the early 1990's with another call for all students to have some degree of ``scientific literacy'' which was approached by the introduction of national and state standards \parencite{Collins}. This came yet again from the idea that every US citizen should be able to engage in scientific dialogue and have an appreciation and understanding about scientific advancement in the modern world \parencite{NAP}. The same rigor framework can be applied to the various national science standards produced throughout the 1990's and 2000's, as well as changes made with the 2002 No Child Left Behind Act and the advent of Common Core curriculum. Of particular interest would be where national or state standards fall on figure \ref{fig:Figure1}.

\subsection{Critical Analysis Of Rigor}
By somewhat narrowing the meaning of ``rigor'', this work has provided one possible tool which can be useful in analyzing the affect of sociopolitical motivation on science education in the United States. This type of analysis can be an insightful process as states adopt the Next Generation Science Standards (NGSS)\footnote{Or likewise develop their own standards}. By considering the sociopolitical backing for the educational standards many US science teachers are required to use, one can gain an understanding of the areas in which rigor could be increased within one's own context.

\section{Conclusions}
The united states has had a nearly continual restructuring of its motivations for teaching science. These reforms of science education can be viewed through a sociopolitical lens of ``why science'', which leads to a better understanding of decisions that impact the level of rigor of US science education. Though not the only factor, it can be shown that sociopolitical views shape what is taught to students, and the importance of two facets of rigor; cognitive demand and depth of content.
\\

This work did not provide a framework to analyze instruction or assessment in terms of rigor, but it would follow that those aspects of science education are similarly affected by the larger social and political beliefs at any present time in history. Though not all encompassing, this work can provide educators; both those who have a hand in making policy, and those who are in the classroom, with a framework in which to analyze and perhaps define what rigor means. This in turn can lead to more thoughtful and purposeful planning and delivery of instruction, and more care to meet specific holistic goals with students. There is a huge swath of literature on what rigor means today which is of interest for further study, including ideas like Project Based Learning and how different educational philosophies from around the world compare in rigor to the US.

\printbibliography
\end{document}